\documentclass[%
 aip,
 amsmath,amssymb,
 reprint,%
]{revtex4-1}

\usepackage{graphicx}
\usepackage{dcolumn}
\usepackage{bm}
\usepackage{xcolor}
\usepackage[utf8]{inputenc}
\usepackage[T1]{fontenc}
\usepackage{mathptmx}
\usepackage{amsmath}
\usepackage{etoolbox}
\usepackage{siunitx}

\makeatletter
\def\@email#1#2{%
 \endgroup
 \patchcmd{\titleblock@produce}
  {\frontmatter@RRAPformat}
  {\frontmatter@RRAPformat{\produce@RRAP{*#1\href{mailto:#2}{#2}}}\frontmatter@RRAPformat}
  {}{}
}%
\makeatother
\begin{document}

\preprint{AIP/123-QED}

\title[]{Angle-resolved spectroscopy of a molecular monolayer}
\title[]{Directed light emission in molecular monolayers on 2D materials via optical interferences}
\title[]{Directed light emission from monolayers on 2D materials via optical interferences}
\author{P. Trofimov}
 \affiliation{Physics Department, Freie Universität Berlin, Germany}
\author{S. Juergensen}%

 \affiliation{Physics Department, Freie Universität Berlin, Germany}
	
\author{A. Dewambrechies Fernández}
 \affiliation{Physics Department, Freie Universität Berlin, Germany}

\author{K. Bolotin}
  \affiliation{Physics Department, Freie Universität Berlin, Germany} 
  
\author{S. Reich}
  \affiliation{Physics Department, Freie Universität Berlin, Germany}
  
\author{H. Seiler*}
  \affiliation{Physics Department, Freie Universität Berlin, Germany}
  
  \email{helene.seiler@fu-berlin.de}

\date{\today}

\begin{abstract}
 Two-dimensional materials provide a rich platform to explore phenomena such as emerging electronic and excitonic states, strong light-matter coupling and new optoelectronic device concepts. The optical response of monolayers is entangled with the substrate on which they are grown or deposited on, often a two-dimensional material itself. Understanding how the properties of the two-dimensional monolayers can be tuned via the substrate is therefore essential. Here we employ angle-resolved reflectivity and photoluminescence spectroscopy on highly ordered molecular monolayers on hexagonal boron nitride (hBN)  to systematically investigate the angle-dependent optical response as a function of the thickness of the hBN flake. We observe that light reflection and emission occur in a strongly directed fashion and that the direction of light reflection and emission is dictated by the hBN flake thickness. Transfer matrix simulations reproduce the experimental data and show that optical interference effects in hBN are at the origin of the angle-dependent optical properties. While our study
focuses on molecular monolayers on hBN, our findings are general
and relevant for any 2D material placed on top of a substrate. Our findings demonstrate the need to carefully choose substrate parameters for a given experimental geometry but also highlight opportunities in applications such as lighting technology where the direction of light emission can be controlled via substrate thickness.

\end{abstract}

\maketitle

\section{\label{sec:level1} Introduction }
In recent years, two-dimensional (2D) monolayers with a wide range of properties - metals, semimetals, topological insulators, semiconductors, insulators, 2D magnets - have been demonstrated \cite{Novoselov2016,Gibertini2019}. 
In the vast majority of cases, such monolayers are grown or deposited on a substrate. This substrate, often a 2D material itself, is much more than a mechanical support. Together, the monolayer and the substrate form a multilayer structure, with widely tunable physical properties via materials' choice or parameters such as the angle between the two components of the heterostructure \cite{Yankowitz2019}. This unprecedented tunability has given rise to exotic electronic states and offers new opportunities for opto-electronic devices. In this context, gaining an understanding of how the substrate affects the optical response of monolayers is essential. One of the simple yet impactful optical phenomena is interference effects arising from multiple light reflections within the substrate \cite{Zhang2015, Kim2018, Lien2015, Li2012, Yoon2009, Buscema2014}. For instance, variation of substrate thickness results in a factor of 10 and 30 of the photoluminescence and Raman intensities modulation in transition metal dichalcogenides \cite{Lien2015} associated with Fabry-Perot resonances. Most previous works so far have focused on the normal incidence case \cite{Zhang2015, Lien2015, Li2012, Yoon2009, Buscema2014}, while the contribution of higher angles to the angle-averaged spectrum was argued to be small. However, interference effects can feature a pronounced angle dependence given the change in optical path lengths for light impinging the sample at different angles \cite{BrotonsGisbert2017, BrotonsGisbert2019, Sigl2022}. Therefore, it is desirable to systematically characterize the angle-dependent optical properties of 2D materials and their substrate parameter dependencies. This is the aim of our manuscript.

To conduct our study, we choose ordered molecular monolayers on hBN of various thicknesses as an examplary heterostructure. Importantly, however, our findings are generally applicable to 2D monolayers on substrates. Before we dive into the study, we first introduce key properties of 2D ordered molecular monolayers to the reader with the salient physics. The Coulomb coupling of transition dipole moments in molecular aggregates leads to the delocalization of excitonic states across multiple molecules. In turn, these states can give rise to remarkable emergent collective phenomena such as superradiance \cite{Dicke1954, Mller2013, Doria2018, Masson2022} and strong light-matter coupling \cite{Hertzog2019}. The study of these phenomena has a rich history, including various types of molecular aggregates, typically classified as J, H or JH types \cite{Deshmukh2022}. Of particular interest are perfectly ordered 1D or 2D molecular lattices, similar to what is achieved in the cold atom community\cite{Rui2020, Srakaew2023} or in supercrystals composed of nanoparticles \cite{Mueller2020, Cherniukh2021}. 

Fabricating well-ordered 2D molecular lattices over micron-scale distances has been a significant challenge. The obtained lattice structure and degree of order depend heavily on substrate parameters, especially surface roughness and defects \cite{Gnder2020, Gnder2023}. Recently, it was discovered that 2D materials such as hBN or graphene act as a template for growing ordered monolayers of 
N,N'-dimethyl-3,4,9,10-perylentetracarboxylicdiimide (MePTCDI) \cite{Zhao2019, Juergensen2023, Zhang2024-ly}. In these structures, due to the high degree of order achieved through the growth process, the situation approaches that of a perfect aggregate. Simulations in prior work by some of us (Ref. \onlinecite{Juergensen2023}) show that a real-space microscopic dipole model captures these observables and yields an exciton delocalized over $\approx$ 20 molecules. We refer to this state as a collective molecular state. These ordered molecular monolayers on 2D materials offer an ideal platform to investigate intermolecular interactions in two dimensions and the collective phenomena mentioned above \cite{Gmez1998, Wrthner2011, Zhao2019, Juergensen2023, Zhang2024-ly}. Furthermore, the physical properties of the delocalized excitonic states may be highly tunable via the choice of the molecule or utilizing substrate properties such as chemical composition, doping level or thickness of the substrate layers \cite{Chaves2020,Waldecker2019}. Understanding how various substrate parameters can be used to control the optical properties of the collective molecular states is therefore essential. For these reasons, this heterostructure provides a perfect test system for the purpose of our study.

\begin{figure*}[ht]
    \includegraphics[]{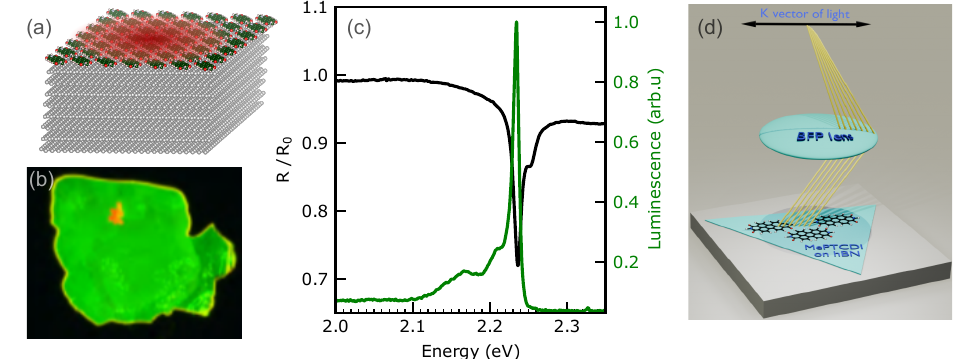}
    \caption{\label{fig1} (a) Illustration of the molecular 2D lattice structure formed by MePTCDI molecules ordered on a hBN substrate. The red halo illustrates the delocalized nature of the exciton hosted in such a structure, i.e. the collective molecular state. (b) Exemplary photoluminescence microscopy image showing homogeneous emission around 2.24 eV. Small remaining aggregate regions (red) demonstrate that monolayer regions can be clearly identified based on the photoluminescence emission wavelength. (c) Exemplary reflectivity (black) and photoluminescence spectra (green) of a MePTCDI molecular monolayer on hBN acquired at 3.5 K. (d) Concept of back focal plane imaging. Ray tracing illustrates how all reflected/emitted light with the same k-vector focuses on a single point on the detector, thereby providing the light-momentum resolution.}
\end{figure*}

In this paper, we study angle-dependent optical properties of MePTCDI monolayers evaporated on top of hBN flakes of various thicknesses using back focal plane imaging spectroscopy. We show that the angular distribution of reflectivity and photoluminescence intensity of the heterostructures widely varies for different hBN flake thicknesses, exhibiting the maximum photoluminescence peak intensity at different incidence angles up to 50 degrees (limited by the numerical aperture of the imaging setup). 
We reproduce our experimental angular reflectivity data using transfer matrix simulations and demonstrate that the observed angular dependence of the reflectivity and photoluminescence signals arise from angle-dependent interference effects within the substrate. The nature of the observed angular dependent  optical response is entirely controlled by the thickness of the hBN flake. While our study focuses on monolayers of MePTCDI, our findings are general and relevant for any 2D material placed on top of a substrate. Our results highlight the importance of optimizing substrate thickness for a particular experiment or application.

\section{\label{sec:level2} Methods }
\subsection{Sample growth}
The sample preparation was carried out by physical vapor deposition (PVD). We mechanically exfoliated hBN flakes of various thicknesses and transferred them onto a 290 nm thick Si/SiO$_2$ substrate. In a tube furnace, the MePTCDI molecules (supplier: TCI) were evaporated and carried to the mechanically exfoliated hBN by an argon gas flow. The growth time was 60\,min and the growth temperature was 230$^\circ$C. The whole process was carried out in a vacuum. The spectroscopy experiments were reproduced on two independently grown sample batches, each batch consisting of several hBN flakes. Overall, we studied MePTCDI monolayers with hBN thickness in the range of 6 - 150 nm. All thicknesses were determined with AFM measurements. 

\subsection{Back focal plane imaging experiments}
An ideal method to investigate angle-dependent optical properties is back focal plane (Fourier) spectroscopy \cite{LeThomas2007}. Instead of rotating the sample with respect to the incident light with a goniometer, the angle resolution is achieved by employing a high numerical aperture objective, see Fig.~\ref{fig1} (d). Back focal plane imaging belongs to the methods of choice to investigate polaritonic band structures \cite{Li2021, Zhang2024-ly, Wertz2010, Dirnberger2023}, but also to reveal out-of-plane dipole emission in materials or heterostructures \cite{Schuller2013, Scott2017, Gao2017, Schneider2020, BrotonsGisbert2017, BrotonsGisbert2019, Han2022, Sigl2022}.
To study the low-temperature excitonic response of MePTCDI, we employ an angle-resolved optical confocal microscopy setup integrated with a closed-cycle Helium Cryostat. The prepared samples were cooled down to 3.5 K and studied in the reflection geometry using a high magnification objective (Zeiss, 100X, NA=0.75). To measure the reflectivity spectra, a stabilized halogen white light source was used, while photoluminescence was excited with a continuous wave laser (532 nm 0.4-2 \textmu W). The spectra were captured using a commercial spectrometer (IsoPlane 320, PIXIS400 CCD camera). Angle-resolved reflectivity spectra were taken using lens imaging of the back-focal plane of the objective in 4F geometry. To obtain reference reflectivity spectra, samples were heated up to 400°C to remove the molecules from the hBN flakes and measured again using the same setup configuration. This is a necessary step to ensure proper normalization of the reflectivity spectra to the hBN background, as PVD growth covers the entire hBN flake. Reflectivity and photoluminescence spectra taken at the sample positions after heating confirmed the successful evaporation of molecules.

\subsection{Simulations}
To simulate the angle-dependent reflectivity spectra from the hBN/MePTCDI sample, we used the PyGTM open-source code (https://github.com/pyMatJ/pyGTM). This Python code is based on the transfer matrix approach and translates a Matlab code originally developed by Passler et al. in Refs. \onlinecite{Passler2017, Passler2020} for anisotropic systems. The simulated heterostructures consisted of a MePTCDI monolayer on hBN crystals of various thicknesses, as measured by our AFM measurements, on top of the Si/SiO$_2$(290 nm) substrate. The hBN flake and the molecular monolayer were modelled as uniaxial materials:
hBN with 
\begin{math}\epsilon_x= \epsilon_y = 4.98\end{math}, 
\begin{math}\epsilon_z=3.03\end{math} 
and the molecular monolayer with 
\begin{math}\epsilon_x= \epsilon_y = 4.95+L_1+L_2\end{math},  
\begin{math}\epsilon_z=1.9\end{math} \cite{Zang1991-lh}. 
In the last expression, L1 and L2 are two Lorentz oscillators corresponding to two observed resonances in the experiment. The values of L1 and L2 were optimized for best fit for the sample with 6 nm hBN. They were kept constant in subsequent simulations, so as to isolate effects from substrate from oscillator strength.
The calculated spectra were normalized to the spectra obtained from the same system but without MePTCDI, matching the experimental normalization procedure. It should be noted that the nonuniformity of the normalized calculated spectra comes from the nonzero epsilon infinity of the MePTCDI layer\cite{Zang1991-lh}.

\section{\label{sec:level3} Results }
\subsection{\label{sec:level31} Overview of sample properties}
The structure of the grown MePTCDI 2D lattice on hBN is shown in Fig.~\ref{fig1} (a). Based on previous high-resolution atomic force microscopy measurements carried out on similar samples (see Ref.\onlinecite{Juergensen2023}, same growth procedure, different batches), the lattice parameters of this structure follow a brick stone lattice with parameters a = b = (11.8 $\pm$ 0.3) \si{\angstrom} and an angle (a, b) = (84 $\pm$ 2)$^{\circ}$. A photoluminescence microscopy image of an exemplary MePTCDI on hBN structure, Fig.~\ref{fig1} (b), displays homogeneous photoluminescence centered at 2.24 eV. As reported in previous works \cite{Juergensen2023, Zhao2019}, this green photoluminescence signal is attributed to the ordered areas of the structure and can be clearly distinguished from the agglomerate regions, which feature a photoluminescence peak in the red at 1.75\,eV (709\,nm). The small red patch visible in Fig.~\ref{fig1} (b) shows an exemplary agglomerate region. Fig.~\ref{fig1} (c) displays reflectivity and photoluminescence spectra obtained at a temperature T = 3.5 K. The reflectivity spectrum of the MePTCDI/hBN structure as a function of photon energy, shown in black, is normalized by the reflectivity of the hBN flake $R_0$. The large reflectivity drop at the main exciton resonance position (2.23-2.24 eV), of around 28$\%$ compared to the background, is a direct consequence of the large oscillator strength of the excitonic transition in the 2D molecular lattice. A secondary feature is observed in the reflectivity spectrum at 2.25-2.26 eV. Although the exact nature of this state is still subject to debate, it is assigned to a separate electronic state as opposed to a shoulder arising from a vibronic progression \cite{Juergensen2023}. The maximum of the photoluminescence spectrum, shown as the green line in Fig.~\ref{fig1} (c), coincides with the reflectivity dip in the R spectrum. These measurements are consistent with absorbance and photoluminescence measurements previously performed on similar structures by some of us, which determined a vanishing Stokes shift in the 2D molecular lattices of Me-PTCDI, in contrast with 60 meV Stokes shift observed for the solvated monomer \cite{Juergensen2023}. Furthermore, the linewidth of the photoluminescence peak is in the range of 8-16 meV at 3.5K depending on the sample, indicating a high-quality excitonic resonance. Overall, our data is consistent with previous measurements on the same or similar heterostructures \cite{Zhao2019, Juergensen2023, Zhang2024-ly}.

The observed large oscillator strength, narrow linewidth, and the absence of Stokes shift all point towards the existence of delocalized excitons typical of molecular aggregates \cite{Juergensen2023, Zhao2019}. Such a delocalized exciton is schematically represented in Fig.~\ref{fig1} (a) as the red halo. So far, the nature of the molecular monolayers growth process keeps the design of MePTCDI-based structures limited to only a few substrates \cite{Juergensen2023, Zhao2019}, with hBN being the only optically transparent one in the spectral region of the delocalized exciton.
Therefore, in the next sections we focus on the angle-dependent optical properties of the excitonic features and systematically investigate how these are determined by the hBN substrate thickness. 

\subsection{\label{sec:level31} Angle-dependent photoluminescence measurements}
\begin{figure}[!h]
    \includegraphics[]{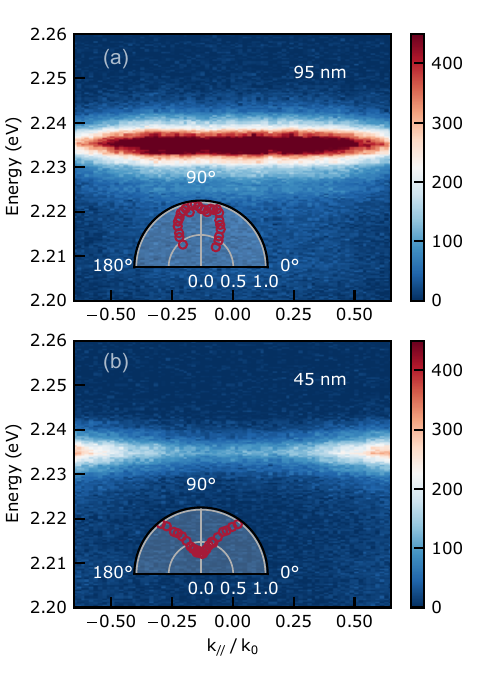}
    \caption{\label{fig-Angle-resolved-PL} (a) Angle-resolved photoluminescence spectra from the MePTCDI structure with (a) 95 nm and (b) 45 nm thick hBN. The insets show a polar plot of the energy cross-section of the angle-resolved spectra taken at the exciton peak position.}
\end{figure}

First, we studied the photoluminescence spectrum as a function of the normalized in-plane wavevector, $k_{\parallel}/k_0$. In Fig.~\ref{fig-Angle-resolved-PL}, we show the angle-resolved photoluminescence signals from two exemplary MePTCDI structures with hBN thicknesses of 95 nm and 45 nm. Signals for additional thicknesses are provided in the Appendix as Fig.~\ref{fig_PL_SI}. We observed a strong dependence of the angular distribution of photoluminescence intensity for hBN flakes of different thicknesses. While the photoluminescence for the structure with 95 nm hBN had maxima at $k_{\parallel}/k_0 = 0$, it had minima at the same angle for the structure with 45 nm hBN (see insets in Fig.~\ref{fig-Angle-resolved-PL}). As a result, for the latter thickness, the observed angle-averaged photoluminescence response of the studied MePTCDI structures is significantly extinguished, especially when low and medium numerical aperture collection optics are used.

\begin{figure}[!h]
    \includegraphics[width=0.45\textwidth]{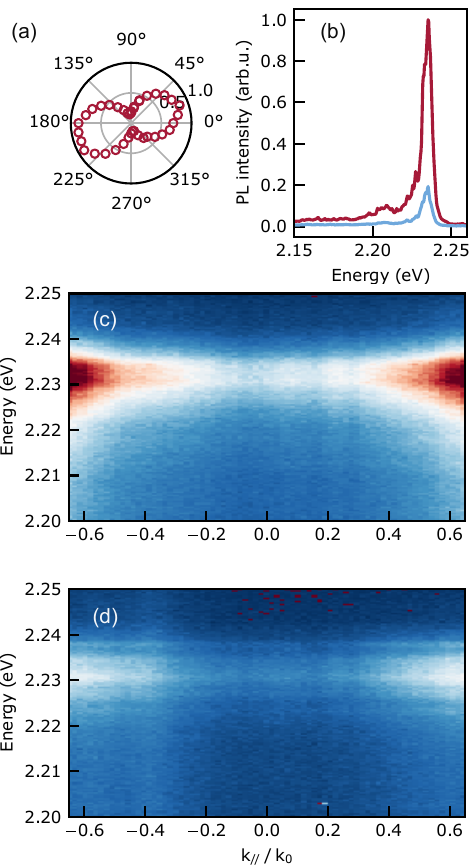}
    \caption{\label{fig-Phi-resolved-PL} (a) Polarization dependence of the photoluminescence peak maxima. (b) Photoluminescence spectra taken at the polarization angle corresponding to the strongest (dark red) and weakest photoluminescence response (light blue). (c-d) Polarization- and angle-resolved photoluminescence for MePTCDI structure with 45 nm hBN. Angle-resolved photoluminescence with the polarizer at (d) 20$^{\circ}$ and (d) 110$^{\circ}$, corresponding to the maximum and minimum of the signal.}
\end{figure}

To study this angle-resolved photoluminescence behaviour in greater detail, we performed polarization-resolved measurements of the same structures with 10$^{\circ}$ rotation step. The obtained spectra had a strongly polarized emission pattern, in line with previous studies \cite{Zhao2019, Juergensen2023}. The results are shown in Fig.~\ref{fig-Phi-resolved-PL}. In panel (a) we report the maximum of the photoluminescence peak as a function of polarizer angle. We note that we observed around 20\% degradation of the signal visible as an asymmetry in the size of two lobes even for the lowest employed of laser power of 1.2 \textmu W. Fig.~\ref{fig-Phi-resolved-PL} (b) shows the average photoluminescence signal at the angles corresponding to minimum (light blue) and maximum (dark red) photoluminescence emission.  
Angle-resolved spectra measured for polarization angle of maxima (Fig.~\ref{fig-Phi-resolved-PL} (c)) and minima (Fig.~\ref{fig-Phi-resolved-PL} (d)) of the photoluminescence features identical angular behaviour.

From the polarization-resolved plot, we can define the linear polarization contrast (\begin{math}P = (I_{max}-I_{min})/(I_{max}+I_{min})\end{math}, where $I_{max}$ and $I_{min}$ are the maximum and minimum intensities. We find $P$ to be around 72\% for two independent samples, a value slightly smaller than that reported in Ref. \onlinecite{Zhao2019} at similar temperatures. 


\subsection{\label{sec:level31} Angle-dependent reflectivity measurements}
To understand the nature of the anisotropic angular dependence of the photoluminescence, we studied the resonant response of the MePTCDI structures. We performed angle-resolved reflectivity measurements on the structures as well as numerical simulations of the data. The experimental reflectivity spectrum as a function of the normalized in-plane wavevector, $k_{\parallel}/k_0$  for two exemplary structures with hBN thicknesses of 6 nm and 115 nm is shown in Fig.~\ref{fig-Angle-resolved-R}(a and b) respectively. Further examples for other thicknesses are provided in the Appendix, Fig.~\ref{fig_Ref_SI}. For all hBN thicknesses, an angle-dependent reflectivity spectrum of the MePTCDI structure is observed. While in the case of the 6 nm flake, the reflectivity dip is maximum at larger angles, it has maximum at $k_{\parallel}/k_0 = 0.00 $ for the 115 nm flake. In some samples, we observed a minor apparent redshift of about 2-3 meV of the minimum of the reflectivity dip at $k_{\parallel}/k_0 = 0.75 $ compared to $k_{\parallel}/k_0 = 0.00 $. We note that since the reflectivity resonance is not symmetrical, a shift of its minimum does not necessarily mean a shift of the peak position. Overall, the variation of signal intensity as a function of angle is the same for reflectivity and photoluminescence of a given sample, i.e. a sample which features a maximum reflectivity dip at $k_{\parallel}/k_0 = 0.75$ will also feature a maximum photoluminescence at that angle. This can be seen in the Appendix in Fig.~\ref{fig_PL_SI} and Fig.~\ref{fig_Ref_SI}.

\begin{figure}[!h]
    \includegraphics[]{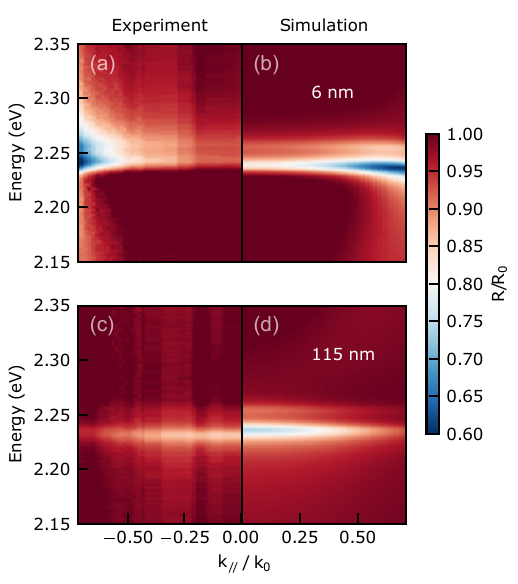}
    \caption{\label{fig-Angle-resolved-R} Angle-resolved s-polarized reflectivity spectra from a monolayer of MePTCDI on a 6 nm (a,b) and 115 nm (c,d) thick hBN flake. (a,c) Experimental reflectivity spectra normalized to the bare hBN background reflectivity R$_0$. (b,d) Simulated angle-resolved s-polarized reflectivity using the transfer matrix method. All plots are shown with the same scale, indicated as the color bar on the right of the figure.}
\end{figure}

\begin{figure}[!h]
    \includegraphics[]{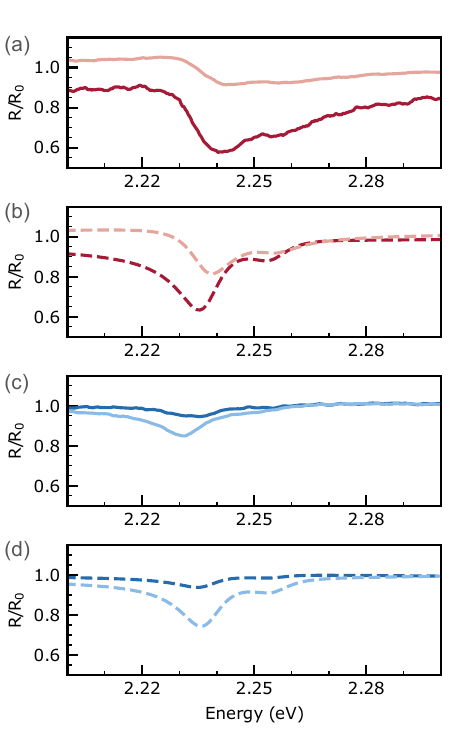}
    \caption{\label{fig-Angle-resolved-R-CS} Angular cross-sections of reflectivity spectra taken from the data shown in Fig.~\ref{fig-Angle-resolved-R} at zero and high incidence angles. Cross-sections of (a) experimental and (b) simulated reflectivity from the structure with the 6 nm hBN taken at $k_{\parallel}/k_0 = 0$ (light pink) and $k_{\parallel}/k_0 = 0.75$ (dark red). Cross-sections of (c) experimental and (d) simulated reflectivity from the structure with the 115 nm hBN taken at $k_{\parallel}/k_0 = 0$ (light blue) and $k_{\parallel}/k_0 = 0.75$ (dark blue).}
\end{figure}

To rationalize these effects, we performed transfer matrix simulations using the open source code PyGTM, see methods section and Refs. \onlinecite{Passler2017, Passler2020}. The simulation results are shown in Fig.~\ref{fig-Angle-resolved-R} (b, d). The comparison of experimental and simulated data in Fig.~\ref{fig-Angle-resolved-R} shows that the simulations qualitatively reproduce the angle-dependent reflectivity signals. Further insights are gained in Fig.~\ref{fig-Angle-resolved-R-CS}, which displays cuts of reflectivity spectra at $k_{\parallel}/k_0 = 0$ and $k_{\parallel}/k_0 = 0.75$ for the same data as in Fig.~\ref{fig-Angle-resolved-R}. Experimental and simulated data are shown in panels (a, c) and (b, d), respectively. Comparing cuts for structures with different thickness of hBN flake (red and blue lines) and different angles of incidence (light and dark colors), it is apparent that the reflectivity peak shape and intensity widely differ. For instance, the reflectivity dip at $k_{\parallel}/k_0 = 0.75$ is only about 5$\%$ in the 115 nm flake, while it is about 15$\%$ for $k_{\parallel}/k_0 = 0$. The trend is qualitatively reproduced by our simulations. The quantitative differences between experimental data and simulations could be due to several factors: (i) the imperfect normalization of our spectra due to the complicated procedure of acquiring reference spectra for this particular molecular heterostructure, and (ii) the oscillator strength variations between different molecular monolayers. We note that for all simulations, we used the same values of the oscillator strengths, but in reality, different samples might have slightly stronger or weaker oscillator strengths.


Therefore, in line with previous results, we observed  \cite{Zhang2015, Kim2018, Lien2015, Li2012, Yoon2009, Buscema2014}, that the interference of light in the underlying substrate, which acts as a Fabry-P\'erot resonator, significantly modifies the reflectivity spectra. More importanly, since the resonance condition of the Fabry-P\'erot cavity depends on the incidence angle, absorption and reflection properties of the structure become largely angle-dependent. In particular when the cavity is tuned out of the resonance and exhibits high reflectivity, the interference of incident and reflected light above the hBN prevents efficient light absorption in the MePTCDI volume. However, when the resonance condition is satisfied, the reflection from the substrate is minimized which results in higher absorption of the overlying monolayer. The shift of the resonance spectral position as a function of substrate thickness leads to a change of its phase at the exciton position. This, in turn, modifies the shape of the excitonic resonance, sometimes leading to changes in the reflectivity dip positions by a few meV (see Fig.~\ref{fig-Angle-resolved-R-CS} (a, b)).

\begin{figure}[!h]
    \includegraphics[width=0.5\textwidth]{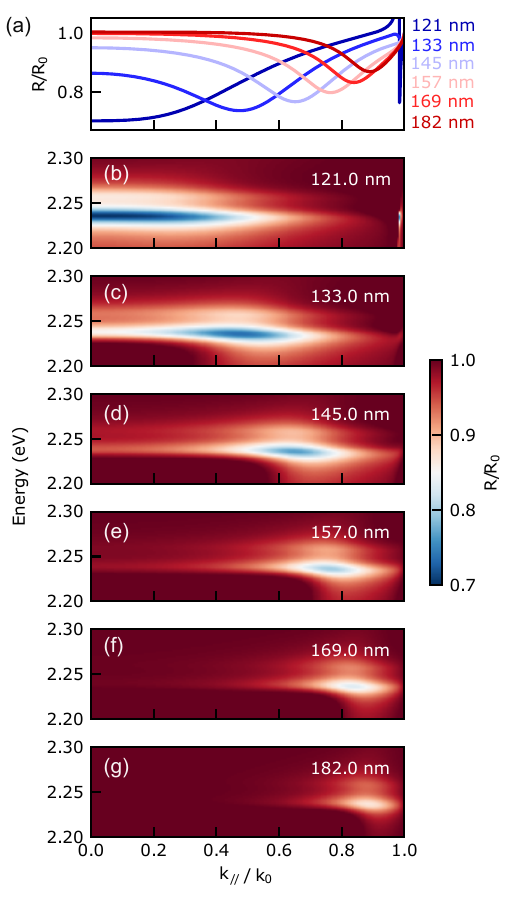}
    \caption{\label{fig-Angle-resolved-R-Sim} Calculated hBN thickness dependence of the p-polarized reflectivity spectra of the MePTCDI structures. (a) Reflectivity spectra at the exciton resonance of MePTCDI structures for a set of different hBN thicknesses (121-182 nm). Thicknesses were chosen to be within half of the period of the Fabry-P\'erot resonance at the exciton energy. (b-g) Energy and angle-resolved normalized reflectivity spectra for the set of different hBN thicknesses corresponding to panel (a).}
\end{figure}

To further demonstrate the significance of angular dependence, we performed transfer matrix simulations of the angle-resolved reflectivity spectra for MePTCDI structures with different hBN thicknesses. Energy cross sections of the angle-resolved spectra taken at the exciton resonance position, shown in Fig.~\ref{fig-Angle-resolved-R-Sim} (a), reveal the angle dependence of the structure's reflectivity dip. The thickness of the hBN layer defines the angle of incident light at which maximum absorption and photoluminescence occur (Fig.~\ref{fig-Angle-resolved-R-Sim} (b-g)). As seen in Fig.~\ref{fig-Angle-resolved-R-Sim}(a), the angular spread of the excitonic feature depends significantly on the thickness of the hBN flake, with thicker flakes yielding less angular spread.

\section{Discussion and summary}
We demonstrated that the angular distribution of reflectivity and photoluminescence intensity of a molecular monolayer is largely controlled by the thickness and the dielectric properties of the underlying substrate. The observed trends are supported by simulations based on the transfer matrix method, which provide insights into the origin of this strongly angle-dependent pattern. Similar to the normal incidence case discussed elsewhere -- which is just a special case of our work, i.e. $k_{\parallel}/k_0 = 0$ -- we explain the angle-dependent signals by optical interference phenomena for the incident and reflected light, which depend both on the dielectric function and thickness of the materials composing the heterostructure. In other words, the Fabry-P\'erot resonances hosted by substrate and hBN flake allow for non-trivial angle dependence of the optical response of the structure.


In summary, the substrate thickness acts as a powerful tuning knob of the optical properties of MePTCDI/hBN heterostructures. Thicknesses should be carefully chosen and optimized for a given experiment, e.g. considering the numerical aperture of objectives used to collect light. These results also provide a detailed analysis of how substrate thickness could be employed for optimizing features in angle-resolved spectra, as sometimes already exploited in polaritonic systems \cite{Li2021}. Directional light emission might be relevant in several photonic technologies such as lasers or light-emitting diodes. One may thus predefine the angle of light emission desired for a given application by optimizing substrate thickness. In the case of MePTCDI, the strong polarization-dependence of the emitted light may act as a further control knob. While the peak width and oscillator strength increases and decreases respectively as temperature rises, the angle-dependent properties remain the same since the interferences are vastly dictated by the underlying flake thickness and its dielectric function, which changes with temperature can be neglected.


Finally, our findings can be generalized beyond MePTCDI on hBN. We modeled the experimental spectra with a generic Lorentz oscillator, and material-specific properties were entered only via the dielectric function. Therefore, a similarly pronounced angle-dependent optical response would be found for other 2D materials on hBN, or other similar substrates capable of supporting Fabry-P\'erot resonances.

\begin{acknowledgments}
This work was supported by the SupraFAB Research Center. S.R. and S.J. acknowledge the Deutsche Forschungsgemeinschaft (DFG) under grant 440298568. H.S. thanks the DFG within Transregio TRR 227 Ultrafast Spin Dynamics (Project B11, Project-ID 328545488).
\end{acknowledgments}

\section*{Data Availability Statement}

Upon final publication of the study, the data used in this manuscript will be posted in an online repository.

\bibliography{aipsamp}

\appendix

\section{Appendixes}
\begin{figure}[!h]
    \includegraphics[]{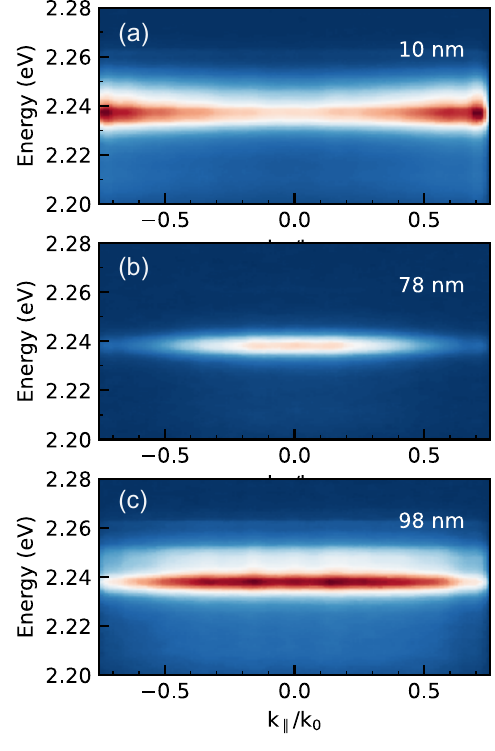}
    \caption{\label{fig_PL_SI} Angle-resolved photoluminescence spectra for additional hBN thicknesses, complementing Fig. \ref{fig-Angle-resolved-PL} of the main text. (a) with a 10-nm thick hBN flake (b) with a 78-nm thick hBN flake and (c) with a 98-nm thick hBN flake.}
\end{figure}

\begin{figure}[!h]
    \includegraphics[]{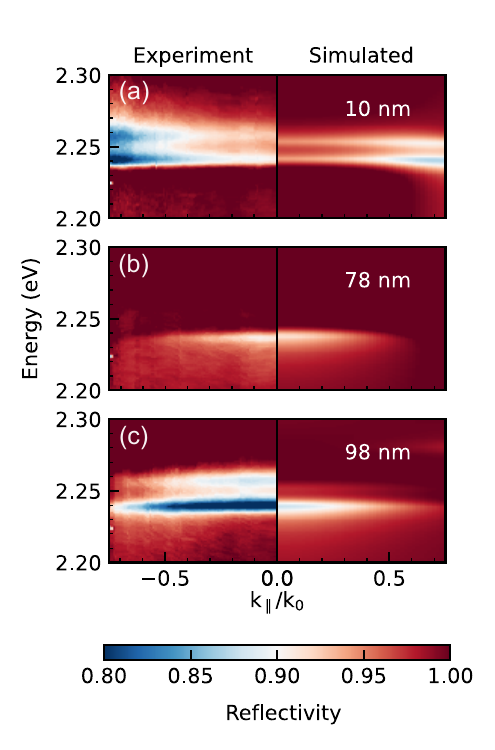}
    \caption{\label{fig_Ref_SI} Angle-resolved reflectivity spectra for additional hBN thicknesses, complementing Fig. \ref{fig-Angle-resolved-R} of the main text. The left part of the figure panels shows the data from experiments, while the right part shows simulated data for (a) a 10-nm thick hBN flake. (b) a 78-nm thick hBN flake and (c) a 98-nm thick hBN flake. We note that here that in contrast to the main text, the normalization procedure is was done as a post-processing step of the data. Here we did not measure reference from bare hBN as was done for the data shown in the main text, but rather used an empirical function. This function fits the reflectivity of the structure around the exciton resonance, with the assumption that there are no pronounced features around it. Using bare hBN flakes is more reliable, however, the empirical background subtraction yields qualitatively correct features, which are well-captured by our simulations. For this series of samples, the oscillator strength in the simulations was optimized to match best the optical response of each sample.}
\end{figure}

Figures \ref{fig_PL_SI} and \ref{fig_Ref_SI} show angle-resolved reflectivity and photoluminescence plots for additional samples.

\end{document}